\documentclass[twocolumn,showpacs,prb,amsmath,amssymb]{revtex4}

\usepackage{graphicx}
\usepackage{bm}

\begin{document}

\title{Metallic phase in a two-dimensional disordered Fermi system with singular
interactions}

\author{Victor M. Galitski}
\affiliation{Kavli Institute for Theoretical
Physics, University of California, Santa Barbara, CA 93106-4030}
\affiliation{Department of Physics, University of Virginia, Charlottesville, VA 22904-4714}

\begin{abstract}
We consider a disordered system of gapless fermions
interacting with a singular transverse $(2+1)$-dimensional gauge-field. We study quantum corrections to fermion
conductivity  and show that they are very different from those in a Fermi liquid with non-singular interactions. 
In particular, the weak-localization effect is suppressed by  magnetic field fluctuations. We argue that these fluctuations 
can be considered static at time scales of  fermionic diffusion. By inducing fluxes through diffusive loops that contribute to weak localization, 
they dephase via the Aharonov-Bohm effect.  It is shown that while the flux-flux correlator  due to thermal fluctuations of magnetic field
is proportional to the area enclosed by the loop, the  correlator due to quantum fluctuations is proportional to the perimeter of the loop.
The possibility of dephasing due to these quasistatic configurations and the corresponding  rates are  discussed.
We also study interaction induced effects and show that perturbation theory contains infrared divergent terms originating
from unscreened magnetic interactions. These singular (Hartree) terms are related to scattering of a fermion off of the static potential created by the
other fermions. We show that due to singular small-angle scattering, the corresponding contributions to the 
density of states and conductivity  are very large and {\em positive} indicating that the fermion-gauge system remains  metallic at low temperatures. 
\end{abstract}


\maketitle

\section{Introduction}
There are many seemingly unrelated physical systems, whose low-energy description
contains fermionic quasiparticles coupled to a fluctuating gauge-field.
A notable example is  a  variety of insulating spin liquid
phases, which are in principle allowed in strongly interacting electronic
systems.~\cite{Lee_etal,Wen_review} These spin liquid states are often described in terms of  non-standard quasiparticles
coupled to dynamic gauge fields.~\cite{Ioffe_Larkin}
Recent experiments of Kanoda~{\em et al.}~\cite{Shimizu_etal,Kurosaki_etal}
indicate  the possibility of a spin-liquid phase in the organic compound $\kappa-(ET)_2Cu_2(CN)_3$.
Motrunich~\cite{Lesik} and S-S. Lee and P.~Lee~\cite{Lee_Lee} argued that the so-called uniform 
resonating-valence-bond (uRVB) phase is the most likely candidate for the phase discovered in these experiments.
This phase is described  by a degenerate Fermi gas of spinons interacting with a dynamic
$U(1)$ gauge field. There are other physical models, whose low-energy description has an identical or very similar
mathematical structure: In particular, the Halperin-Lee-Read  theory~\cite{HLR} of a
compressible Fermi-liquid-like state at the filling factor $\nu = 1/2$ is
formulated in terms of composite fermions interacting with a dynamic 
Chern-Simons  field. 
Another example is a ``vortex metal'' phase suggested in Ref.~[\onlinecite{VG_Fisher}]. 
To access this phase, one starts with a bosonic theory describing preformed Cooper pairs, 
and using duality transforms it into a theory  of bosonic vortices
interacting with a gauge-field. The corresponding dual magnetic field represents the (fluctuating) 
Cooper pair charge density, 
whose mean value can be very large. 
If the dual filling factor is $\nu_{\rm dual} =1/2$, then vortices may become fermions~\cite{Feigel_etal,VG_Fisher} 
forming the Halperin-Lee-Read state (``vortex metal''), which is also described by a  theory with singular
interactions. Remarkably, similar singular  interactions  may  arise 
in more conventional systems in the vicinity of quantum phase transitions, where
soft bosonic fluctuation modes can mimic { longitudinal} gauge-field fluctuations.
Finally we note that strictly speaking physical electrons in metals and semiconductors should be described in terms
of a gauge theory, since they are coupled to a fluctuating [$(3+1)$-dimensional] electromagnetic
field. The corresponding relativistic effects are typically negligible at all reasonable temperatures,
but in the  strict zero-temperature limit, they destroy the usual Fermi liquid picture.~\cite{Reizer} 
An important difference between the usual quantum electrodynamics and  the model 
studied in this paper is the magnitude of the 
effects associated with gauge fluctuations: The corresponding gauge-field-fermion coupling in the artificial
electrodynamics appearing in the context of quantum magnets~\cite{Ioffe_Larkin} or the quantized Hall effect~\cite{HLR} 
is not small, and thus quantum  gauge fluctuations manifest themselves much stronger. 

The aforementioned  fermion-gauge model  has been studied by Nayak and Wilczek~\cite{Nayak_Wilczek}, 
by Altshuler~{\em et al.}~\cite{AIM},
and by Polchinski,~\cite{Polchinski} who  concluded that at least in the limit of small gauge-field-fermion
coupling (large-$N$ limit), it is described by a non-Fermi liquid strong-coupling fixed
point. This fixed point can be visualized as a collection of patches on the Fermi surface,
which do not interact with each other unless the momentum transfer is two Fermi-momenta. 
This phase has a number of other unusual properties: For instance, weak non-singular density-density 
interactions are irrelevant, regardless the sign, and the physics is determined
almost entirely by the interactions between the fermions and  singular transverse 
fields. These results are valid in very clean systems, where the fermion motion is ballistic at 
low-temperatures. 

In this paper we investigate the effects of disorder on the properties
of the fermion-gauge system and study quantum interference effects. 
Understanding these effects of disorder is important in attempts to relate theoretical 
results to real experiments, especially those searching for spin-liquid phases
(in particular thermal transport experiments). 
All real systems  contain finite amount of disorder, which may
strongly affect their properties at low temperatures.~\cite{MIT_review} In a usual Fermi-liquid,
even  weak disorder leads to weak localization~\cite{WL}  and 
similar interaction-induced  effects~\cite{Altshuler_Aronov} and eventually to an insulating behavior.
These quantum interference effects have been observed in numerous experiments and
theoretical results have been quantitatively verified.~\cite{Altshuler_Aronov} An important question
is whether disorder results in a similar localization transition in fermion-gauge systems.
The results presented in this paper  suggest that the fermion-gauge
system may remain metallic at low temperatures, as long as disorder is relatively
weak (dimensionless conductance is large, $\sigma \gg 1$).
This  is due to the existence of  unscreened magnetic forces.
First,  magnetic field fluctuations produce fluxes through closed  
trajectories, which contribute to weak localization. These fluxes result in 
Aharonov-Bohm phase shifts that fermions acquire, while traveling around the diffusive loops
in the opposite directions. This topological effect leads to 
a relatively large dephasing rate at finite temperatures $\tau_\phi^{-1} \sim g^2 \sigma T$
(where $g$ is the dimensionless gauge-field-fermion coupling constant).
We suggest that similar quasistatic configurations due to quantum fluctuations of the magnetic field
may, in principle, lead to a similar dephasing effect with the characteristic rate $\tau_*^{-1} \sim (g^4 / \sigma^2) \tau^{-1}$ 
 (where $\tau$ is the scattering time).
Second, gauge fluctuations lead to interaction corrections of Altshuler-Aronov type.
In a fermion gauge system the leading contributions come from the Hartree processes,
which correspond to the scattering of fermions off of the static interaction potential,
created by the other fermions. Due to the long-range (unscreened) nature of magnetic interactions,
the corresponding scattering rate diverges, due to singular small angle 
scattering. These  Hartree contributions to the density of states and conductivity are {\em positive}, 
which may indicate the stability of the metallic phase.

The structure of the paper is as follows:
 
In Sec.~\ref{Sec-model}, we introduce basic notations and describe the model,
which consists of a degenerate Fermi gas of particles (that we 
occasionally call spinons) coupled to a dynamic non-compact $U(1)$ gauge field
in the presence of disorder. In Sec.~\ref{Sec-model}, we show that 
the clean strong-coupling Nayak-Wilczek fixed point~\cite{Nayak_Wilczek} 
is unstable against weak disorder, and the system should be described by a different diffusive theory.

In Sec.~\ref{Sec-WL}, we qualitatively study the weak localization correction to
conductivity taking into account fluctuating gauge fields. We show that
gauge fluctuations are  slow compared to fermionic
motion, and therefore the magnetic field can be considered static on the time-scales
of fermionic diffusion. We explicitly calculate the correlation function of this
random magnetic field. We argue that dephasing due to this factor is characterized by the flux-flux 
correlator of a typical diffusive trajectory. At finite temperatures, this correlator is proportional
to the area of the path, while at zero temperature the correlator is proportional to the  perimeter of the path. 

In Sec.~\ref{Sec-AA}, we study quantum corrections of Altshuler-Aronov type due to the interactions mediated by the transverse
gauge field. We find that direct perturbation theory contains infrared divergent diagrams. These singular contributions 
to the fermionic self-energy, density of states, and conductivity come from the Hartree
processes in the diffusion channel. The Hartree diagrams and other less divergent exchange diagrams are calculated. 
It is shown that the correction to conductivity is proportional to a formally infrared divergent integral coming from
unscreened magnetic interactions.

\section{The model}
\label{Sec-model}

In this paper we consider a two-dimensional system of fermions coupled to a fluctuating gauge field
in the presence of a disorder potential. The Hamiltonian is 
\begin{equation}
\label{H}
{\cal H} = {\cal H}_{a,f} + {\cal H}_{\rm dis} + {\cal H}_{\rm a},
\end{equation}
where 
{\small
\begin{eqnarray}
\label{H1}
{\cal H}_{a,f} = \sum\limits_{{\bf p},\sigma}
f_{{\bf p},\sigma}^\dagger E({\bf p}) f_{{\bf p},\sigma} + 
\sum\limits_{{\bf p},{\bf q},\sigma}
f_{{\bf p} + {\bf q},\sigma}^\dagger {\bf a}({\bf q}) 
\frac{\partial E({\bf p})}{\partial {\bf p}} f_{{\bf p},\sigma}.
\end{eqnarray}
}
In Eq.~(\ref{H1}), $f^\dagger$ and $f$ are the fermion creation and annihilation operators,
$E({\bf p})$ is the fermionic spectrum, $\sigma = 1,2,\ldots N$ is a spin index,
${\cal H}_{\rm dis}$ is the disorder Hamiltonian, and ${\cal H}_a$ is the bare 
Hamiltonian for the gauge field ${\bf a}$, which can be written as
\begin{equation}
\label{Ha}
{\cal H}_a = \frac{1}{2} \int d^2 {\bf r} \left\{ g_0^2 {\bf e}^2({\bf r}) + 
\frac{1}{g_0^2} \left[ {\bm \nabla} \times {\bf a}({\bf r}) \right]^2 \right\},
\end{equation} 
where ${\bf a}$ and ${\bf e} = - i \frac{\partial}{\partial {\bf a}}$ are canonically  conjugate quantum operators
and $g_0$ is the bare gauge coupling constant.
 In the case of the uRVB phase in the $tJ$-model, 
this bare coupling  is of order~\cite{Nagaosa_Lee} 
$g_0^2  \sim \Delta^2/\rho_{\rm s}^{(0)}$, where $\Delta$ is the charge
gap for holons and $\rho_{\rm s}^{(0)}$ is the bare phase stiffness.

A clean fermion-gauge system ({\em i.e.}, ${\cal H}_{\rm dis} = 0$) has been studied
in Refs.~[\onlinecite{Nayak_Wilczek},~\onlinecite{AIM},~\onlinecite{Polchinski}]
in the large-$N$ limit. It was found that in this limit the system is described
by a non-Fermi liquid strong coupling fixed point, in which the physics is dominated
by  singular magnetic interactions. The results of the large-$N$ treatment 
can be summarized as follows: First, vertex corrections with zero momentum transfer
are small, $\Gamma_0 \sim N^{-1}$. This is due to the fact that gauge bosons 
are slow compared to fermions, and the analog of Migdal's theorem holds.~\cite{AGD} 
Due to  Migdal's theorem, a self-consistent (Eliashberg-like~\cite{Eliashberg}) theory can be developed. In this theory, 
the electronic self-energy depends only on frequency~\cite{Lee,Kim_Lee} $\Sigma(\varepsilon) \propto \varepsilon^{2/3}$,
and is determined by interaction corrections in leading order perturbation theory.
Likewise, the effective action for the gauge field is determined
by fermionic quasiparticles and can be calculated in leading order perturbation
theory. The longitudinal (electric) forces are screened and become effectively
short-range. The transverse (magnetic) forces are not screened, and the corresponding 
gauge-field propagator has the form:
\begin{equation}
\label{clean_propagator}
{\cal K}^{\rm (clean)}_{\alpha\beta}(\omega,{\bf q})  = 
P_{\alpha\beta}^{({\rm tr})} ({\bf q}) \frac{1} {-i\gamma \omega/(v_{\rm F} q) + \chi q^2},
\end{equation}
where $\gamma$ and $\chi$ are constants and $P_{\alpha \beta}({\bf q}) = \delta_{\alpha\beta} - q_\alpha q_\beta/q^2$.
Eventhough, the vertices with small momentum transfers are negligible, the $2 p_{\rm F}$-vertices
are singular. This singularity may in principle lead to a spin-density-wave transition~\cite{AIM2,AILM,Polchinski} 
or a Kohn-Luttinger
pairing instability.~\cite{KL,VG_tbp} Considering energy-scales above the corresponding transition temperatures (if any),
one can study  the following  gapless  fixed point, discovered in Refs.~[\onlinecite{Nayak_Wilczek,AIM}]:
\begin{eqnarray}
\label{Seff_clean}
\nonumber
S = && \int_{\varepsilon,{\bf p}} f_{\varepsilon,{\bf p}}^*
\left[ \Sigma(\varepsilon) + v_{\rm F} p_{\perp} +  \frac{p_{||}^2}{2 m^*} \right]
f_{\varepsilon,{\bf p}}\\
\nonumber
&& \hspace*{-0.5cm} + \int_{\omega,{\bf q}} a\left(\omega,{\bf q}\right) \left[ \chi q^2 + \gamma \frac{|\omega|}{q} \right]
a\left( -\omega, -{\bf q} \right) \\
&& \hspace*{-0.5cm} + g \int_{\varepsilon,{\bf k}} \int_{\omega,{\bf q}}
\left[ 
f_{\varepsilon + \omega, {\bf p} + {\bf q}}^* f_{\varepsilon, {\bf p}} {\bf v}_{\rm F}({\bf q}) {\bf a}(\omega,{\bf q}) 
+ \mbox{h.~c.} \right]. 
\end{eqnarray}
This action describes a collection of patches on the Fermi line, which do not interact with each other. 
In Eq.~(\ref{Seff_clean}), $p_{||}$ and $p_{\perp}$ correspond to the directions tangent
and perpendicular to the Fermi line in the center of the region being considered. One can add 
different operators to the action and consider their relevance under scaling:
$\omega \to s \omega$, $p_{\perp} \to s^{2/3}p_{\perp}$, and $p_{||} \to s^{1/3} p_{||}$.
The gauge-field-fermion coupling [the last term in Eq.~(\ref{Seff_clean})] is {\em exactly
marginal} and can be considered perturbatively with respect to the Gaussian part. 
This allows for controlled calculations within the renormalization group scheme.
Using this scheme, one can find that  non-singular density-density interactions
are irrelevant at the tree level, which justifies dropping  short-range interactions and 
screened electric forces.

We are interested in the effects of disorder on  the properties of the fermion-gauge system.
Let us add the following   term to action~(\ref{Seff_clean})
$$
S_{\rm dis} = \int d^2 {\bf r} V({\bf r}) \int dt f^*(t,{\bf r}) f(t,{\bf r}), 
$$
where $V({\bf r})$ is a quenched disorder potential with the correlation function
$$
\left\langle V({\bf r}) V({\bf r}') \right\rangle = u_0 \delta({\bf r} - {\bf r}').
$$
Using the replica technique, we perform an averaging over disorder realizations and obtain the following
contribution to the action
\begin{equation}
\label{Snn}
\delta S_{nn'} = u_0 \int d^2 {\bf q} {\cal O}_n({\bf q}) {\cal O}_{n'}(-{\bf q}),
\end{equation}
where $n$ and $n'$ are the replica indices and 
${\cal O}_n({\bf q}) = \int_{\varepsilon,{\bf p}} f^*_{\varepsilon,{\bf p} ,n} 
f_{\varepsilon,{\bf p} + {\bf q},n}$. This operator scales as 
\begin{equation}
\label{O}
{\cal O}(s^{2/3} q_{\perp}, s^{1/3} q_{||}) \to s^{-2/3} {\cal O} ({\bf q}).
\end{equation}
From Eqs.~(\ref{Snn}) and (\ref{O}), we conclude that operator (\ref{Snn}) 
is a {\em relevant perturbation}, and therefore the Nayak-Wilczek strong coupling phase 
is unstable against weak quenched disorder, and flows toward a diffusive phase. 

To access this diffusive phase, we will go back to Eq.~(\ref{H}) and include
disorder at the very beginning. The strategy is to treat gauge interactions 
perturbatively with respect to the diffusive Fermi liquid fixed point. An important
issue is to determine a parameter regime ({\em i.e}, temperature and couplings),
in which the perturbative description is reliable. Another important problem is to calculate  
 interaction corrections to the properties of the diffusive Fermi liquid fixed point
in this regime. The main difficulty is due to the fact that we must treat disorder and 
singular gauge interactions self-consistently. 

Our model consists of a diffusive Fermi liquid coupled to a dynamic gauge field.
The propagator of this gauge field is determined by the fermions.
In the diffusive regime,  Landau damping [the term $|\omega|/q$ in Eq.~(\ref{clean_propagator})]
is replaced with the dissipative term $\propto |\omega|$.
 The corresponding propagator of the transverse gauge field 
reads
\begin{equation}
\label{propagator}
{\cal K}^{\rm (diff)}_{\alpha\beta}(\omega,{\bf q}) = 
P_{\alpha\beta}^{({\rm tr})} ({\bf q}) \frac{1} { \sigma |\omega| + \chi q^2},
\end{equation}
where $\sigma = \left( E_{\rm F} \tau \right)/\pi \gg 1$ is the dimensionless
conductance, which is assumed large, $\tau = 1/(2 \pi \nu u_0)$ is the scattering time, 
$\nu = m/\pi$ is the density of states, and $\chi = \delta/m$ is a 
{non-universal} ``Landau diamagnetic susceptibility'' of fermions, which depends strongly on the
form of the fermionic spectrum, and $\delta \sim 1$ is a numerical coefficient characterizing this 
non-universal dependence [in a Fermi gas, $\delta_0 = 1/(12 \pi)$].
We also assume that the dimensionless fermion-gauge-field coupling constant $g$ is small,
which corresponds to the large-$N$ limit with $g \sim N^{-1/2} \ll 1$.
In what follows we will treat the dimensionless fermion-gauge-field coupling as a formal small
parameter without referring to the large-$N$ approximation.  
In this paper, we mostly concentrate on the effects due to  a singular {\em transverse} gauge field,
and assume that all the other interactions (including  dynamically screened
electric forces) are not important. We emphasize that treatment of singular density-density interactions is 
similar. Such singular longitudinal fluctuations may appear in the vicinity of quantum phase transitions.

\section{Dephasing by magnetic field fluctuations}
\label{Sec-WL}
In this section, we qualitatively discuss the weak localization correction to conductivity in
the presence of a fluctuating magnetic field. We assume the validity of the Fermi liquid description for fermions. 
The main equations of this section are the static correlation functions (\ref{K_class_stat}) and (\ref{Kas}),  
the corresponding flux-flux correlators (\ref{FF_class}) and (\ref{FF_quant}), and possible dephasing rates (\ref{t_class}) 
and (\ref{tau_*}).

\vspace*{0.1in}
The weak localization effect originates from constructive interference between closed
diffusive paths. The probability for a diffusive fermion to return to the vicinity 
$d^2{\bf r}$ of the origin in a time $t$ in two dimensions is $C(t) = 1/ \left( 2 \pi D t \right) d^2{\bf r}$.
The contribution of these closed trajectories to conductivity is 
$$
\delta \sigma_{\rm WL} = - \frac{e^2}{2 \pi^2} \int_\tau^{\tau_\phi} \frac{dt}{t}. 
$$
This integral diverges logarithmically if the upper limit is taken to infinity.
In the presence of inelastic processes, the electron coherence is destroyed.
Such processes are characterized by the dephasing time $\tau_\phi(T)$ and the corresponding coherence
length $L_\phi(T) = \sqrt{D \tau_\phi(T)}$. These inelastic processes originate from electron-electron
and electron-phonon scattering events.  The corresponding dephasing rate can be calculated self-consistently 
and in the case of electron-electron interactions is of order $\tau_{\phi}^{-1}(T) \sim \sigma^{-1} T$.
The energy transfer in  these dephasing processes is smaller or of the order of  temperature.~\cite{Aleiner_etal}
At zero temperature and magnetic field, the dephasing time and the weak localization correction diverge. 

In the presence of an external magnetic field ${\bf H}$, the weak localization singularity is cut-off
at time-scales of order $\tau_{\rm H} = 1/(4 e D H)$ even at zero temperature. This suppression is 
due to the  Aharonov-Bohm phase difference that  fermions collect, when traveling around the closed trajectories 
in the opposite directions. In the presence of this time reversal symmetry breaking perturbation ${\bf H}$,
the weak-localization correction remains finite even at zero temperature.

Now let us consider spinons, moving in the presence of a fluctuating gauge field.
Consider a  closed diffusive path $\Gamma(t)$, which among infinitely many others contributes to weak localization
(where $t$ is the time that a fermion needs to travel around the loop).
Let $\Phi\left[\Gamma(t)\right]$ be the flux through the diffusive loop $\Gamma(t)$. This flux depends
on the configuration of the gauge-field. Averaging over all possible gauge-field
realizations leads to a zero mean flux $\left\langle \Phi\left[\Gamma(t)\right] \right\rangle_{\bf a} = 0$.
The crucial observation is that this zero flux has nothing to do with the typical
flux through the loop in a given gauge-field realization. Indeed the average value
of the flux squared is non-zero $\left\langle \Phi\left[\Gamma(t)\right]\Phi\left[\Gamma(t)\right] \right\rangle_{\bf a} \ne 0$.
This finite value leads to Aharonov-Bohm phase shifts and effectively to dephasing. 
We emphasize that this contribution  to dephasing is not related to any inelastic scattering
and has a purely topological nature. In some sense, it is  due to  broken time-reversal
symmetry in each given realization of the gauge field. 
This simple intuitive picture  assumes a static gauge-field background at the time scales
of fermionic diffusion. We argue that this is a very good approximation as the ``photons''
are slow compared to the spinons (at least in the metallic state). Indeed the gauge-field propagator is determined 
by the coupling to matter fields and has the form~(\ref{propagator}): 
$K(\omega,{\bf q}) \propto \left( - i\sigma \omega + \chi q^2 \right)^{-1}$.
This propagator  describes ``diffusive photons'' with the effective diffusion coefficient
$D_a = \chi / \sigma$. In a Fermi liquid, the diamagnetic susceptibility is of the order of
the inverse density of states $\chi \sim 1/m$ and therefore $D_a \sim \sigma^{-2} D$. We see
that as long as the dimensionless conductance is  large, the ``diffusion coefficient''
of photons is much smaller that the diffusion coefficient of fermions. This means that the flux through
a given diffusive trajectory does not significantly change while a fermion is traveling around it.  

We now present a more detailed discussion of the  dephasing rate associated with
 magnetic field fluctuations. This issue was first addressed by
Aronov and W{\"o}lfle~\cite{Aronov_Wolfle} and more recently by W{\"o}lfle.~\cite{Wolfle}
In reviewing the effect of classical thermal fluctuations, we  are going to follow the steps 
used in Ref.~[\onlinecite{Aronov_Wolfle}].  Let us   consider conductivity $\sigma^*$ with respect to a field ${\bf A}$, 
which may or may not be the same as the fluctuating gauge field ${\bf a}$. 
The corresponding  coupling constant ${e_*}^2$  appears only as an overall factor.  

The weak localization correction to conductivity is given by~\cite{Altshuler_Aronov}
\begin{equation}
\label{WL1}
\delta\sigma^*_{WL} = - \frac{4 {e_*}^2 D \tau}{\pi} \int_{-\infty}^{\infty} d \eta\, 
C_{\eta,-\eta}^{t-\eta/2} ({\bf r},{\bf r}),
\end{equation}
where the Cooperon $C$ is the solution of a diffusion-like equation, which in the presence of a classical 
fluctuating field has the form:~\cite{Altshuler_Aronov}
\begin{eqnarray}
\label{Cooperon}
\nonumber
&& \left\{ \frac{\partial}{\partial \eta} + D \left[ -i \frac{\partial}{\partial {\bf r}} 
- g {\bf a} \left(t + \frac{\eta}{2},{\bf r}\right) 
- g {\bf a} \left(t - \frac{\eta}{2},{\bf r}\right) 
\right]^2 \right\} \\
&&\times  C_{\eta,\eta'}^{t} \left[ {\bf r},{\bf r}'; {\bf a} \right] = \frac{1}{\tau} 
\delta(\eta - \eta') \delta({\bf r} - {\bf r}').
\end{eqnarray}
The solution of this equation depends on the realization of the field. To calculate conductivity,
one must find the Cooperon averaged over field fluctuations, 
$C_{\eta,\eta'}^{t} ({\bf r},{\bf r}') = \left\langle C_{\eta,\eta'}^{t} \left[ {\bf r},{\bf r}'; {\bf a} \right]
\right\rangle_{\bf a}$.
%\begin{eqnarray}
%\label{<C>a}
%\nonumber
% C_{\eta,\eta'}^{t} ({\bf r},{\bf r}') = && \!\!\!\!\!  \frac{1}{Z_{\bf a}} \, \int {\cal D} [{\bf a}(x)]
%C_{\eta,\eta'}^{t} \left[ {\bf r},{\bf r}'; {\bf a} \right] \\
%&& \!\!\!\!\!\times \exp\left[ - \int d{x} a_{\alpha}({x})  K_{\alpha \beta}({x})  
%a_{\beta} ({x}) \right],
%\end{eqnarray}
%where $x = (t,{\bf r})$ and $K_{\alpha \beta}(t,{\bf r})$ is the real space-time correlation function of  
%transverse gauge-field fluctuations, which is the Fourier transform of the following  function:
The correlation function of the fluctuating gauge-field is
\begin{equation}
\label{cor_fun}
 K_{\alpha \beta} (\omega,{\bf q}) 
=
 \coth\left(\frac{\omega}{2T}\right) 
{\rm Im} \left[ \frac{ P_{\alpha\beta}^{({\rm tr})} ({\bf q})} {-i\sigma \omega - \epsilon_0 \omega^2 + \chi q^2} \right],
\end{equation}
where $P_{\alpha\beta}^{({\rm tr})} ({\bf q}) = \delta_{\alpha\beta} - q_\alpha q_\beta/q^2$
 and $\epsilon_0 = g_0^{-2}$ is the ``dielectric constant,'' coming from the bare action.
The solution of the diffusion equation for the Cooperon can be formally written as a path integral
over fermionic trajectories. Since the gauge field statistics is Gaussian,  averaging 
over the fluctuations of the magnetic field is straightforward, and we obtain the following
expression:
\begin{eqnarray}
\label{path_int}
\nonumber
C_{\eta,-\eta}^t ({\bf r},{\bf r}') =
\frac{1}{\tau} \!\!\!\!\!
\int\limits_{{\bf r}({\pm \eta}) = {\bf r}} && 
\!\!\!\!\!\!\!\!\!\!{\cal D} [{\bf r}(t)] 
\exp \Biggl\{ -\!  \int\limits_{-\eta}^{\eta} dt_1  
\frac{\dot{\bf r}^2(t_1)}{4 D}\\
&& \!\!\!\!\!\hspace*{-2.4cm}
- \frac{g^2}{2}
 \int\limits_{-\eta}^{\eta} dt_1\int \limits_{-\eta}^{\eta}\! dt_2 
\dot{r}_\alpha(t_1) \tilde{K}_{\alpha\beta}(t_1,t_2) \dot{r}_\beta(t_2) 
\Biggr\},
\end{eqnarray}
In what follows, we will use the notation $\Delta S$ for the last term in
Eq.~(\ref{path_int}). We note that in (\ref{path_int}), $\tilde{K}$ is the sum of the correlation functions:
\begin{eqnarray}
\label{K+K}
\nonumber
\tilde{K}_{\alpha\beta}(t_1,t_2) =&& \!\!\!\!\!\!
K_{\alpha \beta}\left[t_1 + t_2, \Delta{\bf r} \right] +  
K_{\alpha \beta}\left[- t_1 - t_2, \Delta{\bf r} \right]\\
&& \!\!\!\!\!\!\!\!\! \hspace*{-0.3in} + K_{\alpha \beta}\left[t_1 - t_2, \Delta{\bf r} \right] +
K_{\alpha \beta}\left[t_2 - t_1, \Delta{\bf r} \right],
\end{eqnarray} 
with $\Delta{\bf r} = {\bf r}(t_2) - {\bf r}(t_1)$.

At this point, one assumes that the typical frequencies are much smaller than temperature,
which justifies the classical approximation used. This assumption should be verified a posteriori 
by comparing the dephasing rate and temperature. The corresponding classical correlation function
thus reads (we drop here the terms containing higher powers of $\omega$)
\begin{equation}
\label{cor_fun_class}
 K_{\alpha \beta}^{( \rm class )} (\omega,{\bf q}) 
= \frac{ 2 T \sigma}{\sigma^2 \omega^2 + \chi^2 q^4} P_{\alpha\beta}^{({\rm tr})} ({\bf q}).
\end{equation}
To proceed further we make a crucial observation mentioned in the beginning of this section:
The field fluctuations are slow compared to fermions. Really, if $\omega \sim \tau_{\phi}^{-1}$
and $q \sim L_{\phi}^{-1} = 1/\sqrt{D \tau_{\phi}}$, then $\sigma \omega/ (\chi q^2) \sim \sigma^2 \gg 1$
and the static approximation is applicable. In this approximation correlator (\ref{cor_fun_class}) is replaced by
\begin{equation}
\label{K_class_stat}
K_{\alpha \beta}^{( \rm class, static )} ({\bf q}) 
= \frac{ 2 \pi T }{\chi q^2} P_{\alpha\beta}^{({\rm tr})} ({\bf q}).
\end{equation}
This correlation function describes a static random magnetic field with the correlation
function $\left\langle b({\bf r}) b({\bf r}') \right\rangle = (T/\chi) \delta({\bf r} - {\bf r}')$. 
In this case the term $\Delta S = 2 \left\langle \Phi(\Gamma) \Phi(\Gamma) \right\rangle$ is just the same-time 
flux-flux correlator of the quasistatic random magnetic field, arising from thermal gauge fluctuations.
The flux-flux correlator depends on the path and according to  (\ref{K_class_stat}) is proportional to
the area enclosed by the path
$$
\left\langle \Phi(\Gamma) \Phi(\Gamma) \right\rangle_{\rm classical} =  g^2 \frac{T}{\chi} A[\Gamma].
$$
Since the typical area of the trajectory is of order $\overline{A[\Gamma(t)]} \sim D t$ we find
\begin{equation}
\label{FF_class}
\overline{\Delta S} \sim \frac{ g^2 D}{\chi} T t.
\end{equation}
At this qualitative level, one can identify the dephasing rate as 
\begin{equation}
\label{t_class}
\tau_\phi^{-1}(T) \sim g^2 \sigma T.
\end{equation}
It is possible to calculate the corresponding weakly-temperature dependent coefficient, but this
is beyond the scope of the present discussion [according to W{\"o}lfle,~\cite{Wolfle}
$\overline{\Delta S} = { g^2 D}/({8 \pi \chi}) T t \ln{(t/\tau)}$]. 

There are a few important points, which make dephasing by magnetic field
fluctuations drastically different from dephasing by electric field fluctuations:
First of all, the physical mechanism involved in the former case is not related
to any inelastic scattering and has a purely topological origin. Magnetic field
fluctuations are slowed down by fermions and become quasistatic. The fermions see 
the corresponding random background and this leads to dephasing due to a finite flux
through the diffusive loops around which the fermions travel. 
Second, the typical temperature-dependent dephasing rate is proportional to the conductivity
$\tau_\phi(T) T \sim \sigma^{-1} g^{-2}$ [while in the two-dimensional electronic problem,
the situation is exactly the opposite, the  dephasing rate is proportional to the resistivity
$\tau_\phi^{\rm Coulomb} (T) T \sim \sigma$].
We note that if the gauge-field-fermion coupling is not small, then the dephasing rate is much larger
than temperature, which invalidates the assumption about only classical fluctuations
being important. This observation suggests that modes with frequencies larger than temperature may play an important role.

We now consider quantum fluctuations of the magnetic field. We will show that (i)~Quantum fluctuations
of magnetic field can also be considered quasistatic at time scales of fermionic diffusion and (ii)~
Flux-flux correlator  due to these quasistatic fluctuations is proportional to the
{\em perimeter} of the loop.

First, let us calculate the correlation function in  real space-time. 
At zero temperature, this can be done by a straightforward evaluation 
of the Fourier transform of (\ref{cor_fun}). The  result is
\begin{eqnarray}
\label{K12}
\nonumber
K_{\alpha \beta} (x) = &&  
\!\! \left[ K^{(1)} (x) - K^{(2)} (x) \right] 
\left( \delta_{\alpha \beta}
- \frac{r_\alpha r_\beta}{r^2} \right) \\
&& + K^{(2)} (x) \frac{r_\alpha r_\beta}{r^2},
\end{eqnarray}
where
\begin{eqnarray}
\label{K1}
K^{(1)} (x) = 
\frac{1}{8 \pi^2 \chi t} \left[ e^{-u} {\rm Ei} (u) -
 e^u {\rm Ei} (-u) \right]
\end{eqnarray}
and
\begin{eqnarray}
\label{K2}
\nonumber
\hspace*{-0.1in}K^{(2)} (x) = 
\frac{1}{4 \pi^2 \sigma r^2} \Biggl\{ &&
2 \ln\left( \gamma u \right) \Biggr. \\
&& - \left[ e^{-u} {\rm Ei} (u) +
 e^u {\rm Ei} (-u) \right] \Biggr\}.
\end{eqnarray}
In Eqs.~(\ref{K12}), (\ref{K1}), and (\ref{K2}), $x=(t,{\bf r})$ is a short-hand notation for the space-time coordinate,
\begin{equation}
\label{u}
u =  \frac{\sigma r^2}{4 \chi (t + \omega_{\rm m}^{-1})},
\end{equation}
$\gamma \approx 1,781$ is the Euler's constant, and ${\rm Ei}(u)=- \int_u^{\infty} (e^x / x) dx$ is the exponential
integral function. We have introduced a ``high-energy cut-off'' $\omega_{\rm m}$ in the definition of the
parameter $u$ in Eq.~(\ref{u}). This cut-off is needed to regularize the behavior of the correlation function in the $t \to 0$ limit. 
We note that neglecting the term $\epsilon_0 \omega^2$ in Eq.~(\ref{cor_fun}) would lead to a logarithmically divergent equal-time
Fourier transform. This divergence is cut-off by the quadratic part $\epsilon_0 \omega^2$.
We remind that this term comes from the bare action for the gauge field
(in the case of the uRVB state in the $tJ$-model, the  constant $\epsilon_0$ is of the order of
the tunneling amplitude divided by the holon charge gap squared~\cite{Nagaosa_Lee}). 
Since the diffusion approximation breaks down at the time-scales smaller than the scattering time $\tau$,
we can write the cut-off as 
\begin{equation}
\label{wm}
\omega_{\rm m} = {\rm min}\, \left[\sigma/\epsilon_0,\, \tau^{-1} \right].
\end{equation}
We note that results (\ref{K12}), (\ref{K1}), and (\ref{K2}) are exact as long as $t$ is finite
and have  logarithmic accuracy if $t \to 0$. 

Let us now consider fermions diffusing in the fluctuating background described by Eqs.~(\ref{K12}), (\ref{K1}), and (\ref{K2}). 
Since for a typical fermionic diffusive trajectory $|{\bf r}(t)| \sim \sqrt{D t}$, the  value of the parameter $u$ [see Eq.~(\ref{u})] can be estimated as 
$u \sim \sigma D/ (4 \chi) \sim \sigma^2 \gg 1$. We see that this parameter is  large
(as long as the system is metallic), and we can use the asymptotic expression for the correlation functions:
\begin{eqnarray}
\label{Kas}
\nonumber
\lim\limits_{u \to \infty} K_{\alpha\beta}(t,{\bf r}) = 
-\frac{1}{2 \pi^2 \sigma r^2} \Biggl\{&& \!\!\!\!\! \ln \left[ 
\frac{\gamma \omega_{\rm m} \sigma r^2}{4 \chi {\rm e}^2} \right]
\left( \delta_{\alpha\beta} - \frac{r_\alpha r_\beta}{r^2} \right)\\
&& \!\!\!\!\! - \ln \left[ 
\frac{\gamma \omega_{\rm m} \sigma r^2}{4 \chi}\right] \frac{r_\alpha r_\beta}{r^2}
\Biggr\},
\end{eqnarray}
where ${\rm e} \approx 2.718$ is the base of the natural logarithm.
Apparently, Eq.~(\ref{Kas}) describes a static random magnetic field
described by the non-trivial spatial correlation function (\ref{Kas}).

Using correlation function (\ref{Kas}) one can calculate the flux-flux correlator
$\left\langle \Phi \Phi \right\rangle$. 
\begin{equation}
\label{W}
\left\langle \Phi(\Gamma) \Phi(\Gamma) \right\rangle =  g^2 \oint_\Gamma \oint_\Gamma
dr_{1\alpha} dr_{2\beta} \left\langle
a_{\alpha}(t,{\bf r}_1) a_{\beta}(t,{\bf r}_2) \right\rangle.
\end{equation}
The exact value of the flux-flux correlator $\left\langle \Phi(\Gamma) \Phi(\Gamma) \right\rangle$ 
depends on the geometry of the path $\Gamma$ (diffusive paths should be  
averaged over the distances of order mean free path $l$ and viewed as continuous curves).
As an example we can consider a circular path $S^1$ of radius $R$. We have to calculate integrals
of the following type:
$$
I\left[ S^1 \right] = \oint_{S^1} \oint_{S^1}
\frac{d{\bf r}_1 d{\bf r}_2}
{\left| {\bf r}_1 - {\bf r}_2 \right|^2}. 
$$
As it stands, the integral is  divergent at small distances. If we require that 
$\left| {\bf r}_1 - {\bf r}_2 \right|<r_{\rm min}$, we obtain
$$
I\left[ S^1 \right] = \frac{2 \pi R}{r_{\rm min}} - 2 \pi^2.
$$

We see that the leading contribution is proportional to the circumference of the circle, with the
second term being a small correction. This result is universal and applies for any loop:
Indeed, from Eq.~(\ref{Kas}) it follows that  the main contribution 
to the correlator  always comes from the regions, where points ${\bf r}_1$ and ${\bf r}_2$ are 
located very close to each other, regardless the form of the trajectory. This is due to the overall factor $r^{-2}$
in Eq.~(\ref{Kas}). Therefore, for a generic continuous path $\Gamma$, the integral is proportional to the perimeter
of the path, $P[\Gamma]$:
$$
I\left[ \Gamma \right] \propto \frac{P \left[\Gamma \right]} {r_{\rm min}} 
+ \Biggl\{ \mbox{geometrical  factor\,}\sim\,O(1)\Biggr\},
$$
In the context of a diffusive Fermi system, the cut-off distance is of order mean free path
$r_{\rm min} = l$, which is the length-scale at which the diffusive description breaks down.
Within logarithmic accuracy, we can put $r = l$ in the arguments of the logarithms in Eq.~(\ref{Kas}).
Assuming that $\omega_{\rm m} \sim \tau^{-1}$, we find the flux-flux correlator
\begin{equation}
\label{FF_quant}
\left\langle \Phi(\Gamma)\Phi(\Gamma)\right\rangle_{\rm quantum} \sim \frac{g^2 \ln{\sigma}}{\sigma l} {P[\Gamma]}.
\end{equation}
This behavior of the flux-flux correlator can be called a perimeter law. It should be compared with the
``area law'' behavior of the flux-flux correlator of classical thermal fluctuations (\ref{FF_class}).
The perimeter of a typical fermionic trajectory is $P[\Gamma(t)] \sim 2 \pi \sqrt{Dt}$ and therefore the flux-flux
correlation function of a trajectory of this size is $\left\langle \Phi \Phi \right\rangle \sim \frac{g^2 \ln{\sigma}}{\sigma l} \sqrt{Dt}$.
It is plausible to assume that these fluxes may lead to dephasing with  the corresponding time-scale 
being
\begin{equation}
\label{tau_*}
\tau_*^{-1} \sim g^4 \frac{\ln^2{\sigma}}{\tau \sigma^2}. 
\end{equation}

In the usual electronic systems it is known that quantum fluctuations of electric field ({\em i.e.}, longitudinal modes with
$\omega > T$) do not contribute to dephasing in the following sense:~\cite{Aleiner_etal} One can separate different contributions 
to conductivity by their magnetic field dependence (the external magnetic field is assumed to provide the leading
contribution to dephasing with the interaction terms being small corrections). The weak localization term is defined as the leading term
in expansion in $1/\sigma$, which depends on the external magnetic field as $1/B^2$. Diagrammatically it is described
by the maximally crossed diagram. The perturbative statement about the absence of quantum dephasing is that  the first order 
interaction correction 
to the weak localization term does not contain contributions from fluctuation modes with $\omega \gg T$ 
(or more precisely, this contribution is exponentially small).
The corresponding cancellation of the quantum terms is related to the detailed balance in equilibrium. 
The remaining linear-in-$T$ dephasing term is due to inelastic scattering of electrons; the corresponding
rate vanishes at zero temperature.

An interesting (open) question is whether a similar cancellation of ``quantum dephasing terms'' in the maximally
crossed diagram happens in the case of a Fermi gas coupled to a slow fluctuating magnetic field. At the moment, the existence of such terms
would not contradict to any known results. As we have seen above, the main contribution to 
dephasing  is not related to any inelastic scattering, and therefore the detailed balance argument is perhaps
not relevant here. Second, quantum dephasing expected from the above arguments based on the ``perimeter law'' (\ref{FF_quant})
should appear only in second order of a perturbation theory and there are no known analytic results in this order. 
Third, the usual structure of perturbation theory with an external magnetic field providing the leading contribution to the 
weak localization cut-off is not appropriate  in this context, since the main effect, which is based on magnetic field fluctuations, 
will be smeared out by such a large external field. 
From the qualitative arguments described in this section, it seems conceivable that quantum fluctuations of magnetic field may
still lead to a dephasing effect with the characteristic time-scale given by Eq.~(\ref{tau_*}).

\section{Altshuler-Aronov corrections}
\label{Sec-AA}
In this section, we calculate the exchange and Hartree contributions to
the self-energy, density of states, and conductivity. We show that the direct leading order perturbation theory
is ill-behaved since certain diagrams are infrared divergent. These singular contributions
come from the Hartree-type processes. The corresponding corrections to the density of states~(\ref{nu4}) and  
conductivity~(\ref{sigma_H}) are positive.

\vspace*{0.1in}

\subsection{Interaction corrections to the density of states} 
\label{Sub:nu}

In two-dimensional electron liquids, long-range Coulomb interactions produce singular contributions
to the density of states and  conductivity, which typically have the same sign and order of magnitude
that the weak localization correction. 
%(the latter comes from a quantum correction to the diffusion coefficient). 
These Altshuler-Aronov corrections originate from  Coulomb interactions, but in certain
regimes (dense electron liquid) do not depend on the strength of these 
interactions. There is no contradiction here, since in deriving
these corrections one assumes that the temperature is smaller than interactions in the following sense: 
$T/ E_{\rm F} \ll r_{\rm s}^2 \sigma$, with $\sigma = (E_{\rm F} \tau)/\pi$ being the dimensionless conductance
and $r_{\rm s} = e^2/(\hbar v_{\rm F})$. In this parameter regime, the leading correction to the
conductivity is universal and negative. This universality is specific to screened Coulomb interactions.
In the case of a fermion-gauge system, the interaction-induced corrections are due to  gauge fields.
The corresponding forces are unscreened, which leads to a different structure of perturbation theory.
The existence of corrections of this type was emphasized   by Halperin, Lee, and Read~\cite{HLR} in 
the context of the $\nu=1/2$ quantized Hall state. Also, Khveshchenko~\cite{Khveshchenko} and Mirlin 
and W{\"o}lfle~\cite{Mirlin_Wolfle} studied quantum corrections to conductivity, focusing on the exchange (Fock) contributions.

Below we show that the Altshuler-Aronov corrections to the density of states and 
conductivity are dominated by {\em positive Hartree terms}. The corresponding diagrams are found
logarithmically divergent. We suggest that the dominance of positive Hartree terms  correspond
to antilocalization at the one-loop level. The underlying physical argument behind this result is the following:
When a fermion travels through a disordered system,
it sees a static self-consistent potential created by the other fermions. The fermion scatters off of this static
background and this scattering is responsible for Hartree contributions to the density of states and conductivity. 
We note that the static gauge propagator is proportional to $1/q^2$. This kind of unscreened interaction leads
to a divergent single-particle scattering rate (naively, one would expect a power law divergence coming from  
singular small-angle scattering $\left\langle |{\bf p} - {\bf p}'| \right\rangle_{{\bf p}, {\bf p}'}$,
but we show that the divergence is much weaker). Therefore, the effect of scattering of a given fermion off
 of the static background created by the other fermions lead to singular Hartree terms, which
thus dominate the physics. From the theory of electron-electron interactions in disordered conductors, 
it is known that the Hartree contribution to conductivity is 
positive for repulsive interactions. In a fermion-gauge system, the magnetic interactions
are indeed repulsive and unscreened (at least in the non-compact formulation), and thus
the leading order interaction correction to conductivity is expected to be  large and positive.

We also note that a similar situation may occur in the vicinity of quantum phase
transitions in fermionic systems, where the fermions are coupled to a soft bosonic fluctuation mode,
which is in a sense equivalent to a {\em longitudinal} singular gauge field. Indeed, the fluctuation 
propagator in this case is  $K_{\rm fluc}(\omega,q) \propto \left[ -i c \omega/(-i\omega + Dq^2) + q^2 + \xi^{-2}  
\right]^{-1}$, where $c$ is a constant and $\xi$ is the correlation length. The correlation length diverges at criticality,
and the fluctuation mode becomes massless (or unscreened in our language). This should lead to an enhancement
of interaction corrections. 

\begin{figure}[htbp]
\centering
\includegraphics[width=3in]{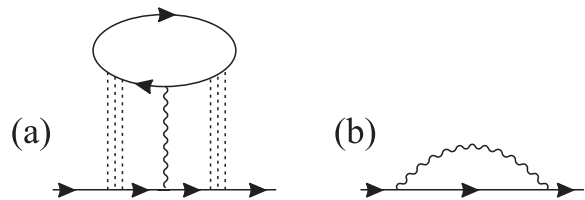}
\caption{Hartree and Fock contributions to the fermion self-energy.
The wavy line corresponds to singular magnetic interactions.}
\label{Self-energy}
\end{figure}

\begin{figure}[htbp]
\centering
\includegraphics[width=3in]{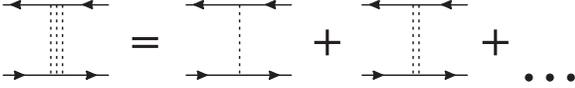}
\caption{Diagram series for diffuson; impurity ladder in the particle-hole channel}
\label{Diffuson}
\end{figure}

The quantum correction to the fermionic self-energy is described by the diagrams presented in 
Fig.~\ref{Self-energy}.~\cite{Lee_Ramak}
In what follows we assume a white-noise correlated disorder potential [see Eq.~(\ref{<uu>})], and therefore 
the vector vertices are not renormalized by disorder. 
The basic ingredients needed to calculate the diagrams in Fig.~\ref{Self-energy} are as follows:~\cite{AGD}
The fermionic Matsubara Green's function
\begin{equation}
\label{G0}
{\cal G}^{(0)}_\varepsilon ({\bf p}) = \frac{1}{i\varepsilon - \xi_{\bf p} + i {\rm sgn}\,\varepsilon/(2 \tau)},
\end{equation}
where $\xi_{\bf p} = E({\bf p}) - E_{\rm F} \approx v_{\rm F} \left(p - p_{\rm F}\right)$.
The impurity propagator (a single dashed line in Fig.~\ref{Diffuson}) reads
\begin{equation}
\label{<uu>}
\left\langle u({\bf r}) u({\bf r}') \right\rangle = \frac{1}{2 \pi \nu \tau} \delta\left({\bf r} -{\bf r}'\right),
\end{equation}
where $\nu = m/\pi$ is the density of states.
We note that the impurity lines do not depend on frequency.
The particle-hole propagator (diffuson) shown in Fig.~\ref{Diffuson} reads 
\begin{equation}
\label{D0}
{\cal D}_{\varepsilon,\varepsilon'} ({\bf p},{\bf p}') = \frac{1}{2 \pi \nu \tau} + \frac{1}{2 \pi \nu \tau^2}
\frac{\theta\left(- \varepsilon \varepsilon' \right)}
{\left| \varepsilon - \varepsilon' \right| + D \left| {\bf p} - {\bf p}' \right|^2}, 
\end{equation} 
where $D = v_{\rm F}^2 \tau/2$ is the diffusion coefficient.

The wavy line represents magnetic  interactions,
\begin{equation}
\label{Kmat}
{\cal K}_{\alpha\beta} (\Omega,{\bf Q}) = P_{\alpha\beta}^{\rm (tr)} ({\bf Q}) \frac{1}{\sigma \left| \Omega \right| + \chi Q^2},
\end{equation}
where $\sigma = ( E_{\rm F} \tau )/\pi$ is the dimensionless conductance and $\chi = \delta/ m$ is the ``diamagnetic
susceptibility'' with $\delta \sim 1$. 
Finally, the points connecting the wavy lines and the solid fermion lines are vector vertices $g {\bf v}$, 
where $g \ll 1$ is the dimensionless coupling to the fluctuating gauge field and ${\bf v} = {\bf p}/m$ is the fermion velocity.

 Using  Eqs.~(\ref{G0}), (\ref{<uu>}), (\ref{D0}), and (\ref{Kmat}), 
one can calculate the fermionic self-energy to the leading order in the
magnetic interaction. The self-energy is the sum of two contributions:
$$
\Sigma = \Sigma_{\rm H}  + \Sigma_{\rm F},
$$
where the first term is the Hartree correction in the diffusion channel (Fig.~\ref{Self-energy}a) 
and the second term is the Fock correction (Fig.~\ref{Self-energy}b).
We do not consider here  effects in the Cooper channel 
(see also Sec.~\ref{Sec-WL}). The Fock term is  relatively small [see Eq.~(\ref{sf}) below], since the vector
vertices are not renormalized by disorder. 
Let us study the Hartree term in the diffusion channel. The corresponding self-energy
has the form:
\begin{eqnarray}
\label{sh}
\nonumber
\Sigma_{\rm H} \left( \varepsilon,{\bf p} \right) =
2 T \sum\limits_{\omega} \int && \!\!\!\!\! \frac{d^2{\bf q}}{\left( 2 \pi \right)^2}
{\cal G}_{\varepsilon + \omega} \left( {\bf p} + {\bf q} \right) \\
&&\!\!\!\!\!\!\!\!\!\!  \hspace*{-0.3in} \times  {\cal D}^2_{\varepsilon,\varepsilon + \omega} \left({\bf p}, {\bf p} + {\bf q} \right) 
B_{{\varepsilon},{\bf p}} \left(\omega, {\bf q}\right),
\end{eqnarray}
where $2$ is the spin degeneracy factor and $B$ corresponds to the block of four
Green's functions (see also Fig.~\ref{Saw-diagrams}c):
 \begin{eqnarray}
\label{B}
\nonumber
 B_{{\varepsilon},{\bf p}} \left(\omega, {\bf q}\right) =
g^2 \int && \!\!\!\!\! 
\frac{d^2{\bf p}'}{\left( 2 \pi \right)^2} \frac{d^2{\bf p}''}{\left( 2 \pi \right)^2} 
v'_\alpha v''_\beta {\cal K}_{\alpha \beta} \left(0, {\bf p}' - {\bf p}'' \right) \\
 && \!\!\!\!\!\!\!\!\!\!\!\!\!\!\!\!\!\!\!\!\!\!\!\!\!\!\!\!\!\!  \hspace*{-0.4in}
\times\, {\cal G}_{\varepsilon} \left( {\bf p}' \right)
{\cal G}_{\varepsilon + \omega} \left( {\bf p}' + {\bf q} \right) 
{\cal G}_{\varepsilon} \left( {\bf p}''  \right) 
{\cal G}_{\varepsilon + \omega} \left( {\bf p}'' + {\bf q} \right). 
\end{eqnarray}
The crucial point is that the interaction line in this block carries no frequency. 
The integral above is infrared divergent due to  long-range magnetic interactions. 
At this point, let us introduce  a small-${Q}$
cut-off $Q_{\rm min}$ as a phenomenological parameter. The value of this parameter can 
not be determined  within the general formulation.
We note that we can not replace the magnetic interaction by its average over the Fermi surface,
since the Fermi-momenta ${\bf p}$ and ${\bf p}'$ are constrained to be very close to each other
due to the dominance of small-angle scattering. Having this in mind, we calculate the $B$-block (\ref{B}), which 
to the leading order is just a constant
\begin{equation}
\label{B0}
B = \frac{g^2 \nu v_{\rm F}^2 \tau^3}{\chi} L,
\end{equation}
where $L = \ln \left( p_{\rm F}/Q_{\rm min} \right)$ is the logarithmic divergence coming from  long-range
interactions. 
Using the self-energy (\ref{sh}) and Eq.~(\ref{B0}), we can calculate the density of states,
\begin{equation}
\label{nu1}
\delta \nu = - \frac{2}{\pi} \int \frac{d^2{\bf p}}{\left( 2 \pi \right)^2} 
{\rm Im}\, \left\{ 
G^2 \left( \varepsilon, {\bf p} \right)  \Sigma \left( i \varepsilon_n \to \varepsilon, {\bf p} \right)
\right\},
\end{equation}
where $G(\varepsilon,{\bf p})$ and $\Sigma \left( i \varepsilon_n \to \varepsilon, {\bf p} \right)$ are the
retarded Green's function and self-energy correspondingly. After the analytical continuation in Eq.~(\ref{nu1}), we obtain
 the density of states
\begin{eqnarray}
\label{nu2}
\delta \nu_{\rm H} (\varepsilon,T) = \frac{ 2 B \nu}{\left( 2 \pi \nu \tau \right)^2}\,
{\rm Im}\, \int && \!\!\!\!\! \frac{d\omega}{2 \pi} \frac{d^2{\bf q}}{\left( 2 \pi \right)^2} 
\frac{1}{\left( - i \omega + D q^2 \right)^2} 
\nonumber \\
&& \!\!\!\!\!\!\!\!\!\!\!\!\!\!\!\!\!\!\!\!\!\!\!\! \!\!\!\!\!\!\!\!  \hspace*{-0.4in}
\times \left[ \tanh \left( \frac{\varepsilon + \omega}{2T} \right) + 
 \tanh\left( \frac{ \omega - \varepsilon}{2T} \right) \right].
\end{eqnarray}
Evaluating the integral at $\varepsilon = 0$ and using the Fermi-gas value of the diamagnetic susceptibility
$\chi = \delta_0/m = 1/(12 \pi m)$ (the result should be multiplied by $\delta_0/\delta$ if the
susceptibility is different), we obtain  the Hartree correction
\begin{eqnarray}
\label{nu3}
\delta \nu_{\rm H} (0,T) = \nu  \frac{3 g^2 L}{\pi^2} \ln\frac{1}{T\tau},
\end{eqnarray}
where $\nu = m/\pi$ and we remind that $L = \int_{Q_{\rm min}}^{p_{\rm F}} dQ/Q \gg 1$ is the
divergent part coming from long-range magnetic interactions. 

The Fock (exchange) correction to the self-energy has the following form:
\begin{equation}
\label{Se}
\Sigma_{\rm F} \left(\varepsilon,{\bf p}\right) = -g^2 T \sum\limits_{\Omega}
\int \frac{d^2{\bf Q}}{\left( 2 \pi \right)^2} 
\frac{p_\alpha}{m} \frac{p_\beta}{m}
{\cal K}_{\alpha \beta} (\Omega,{\bf Q}) {\cal G}_{\varepsilon + \Omega} ({\bf p} + {\bf Q}).
\end{equation}
It was emphasized earlier that this expression does not contain impurity vertices due
to interactions being current-current. It is also clear that this diagram is free of any
infrared divergences.
The corresponding integrals can be easily calculated and we obtain the following expression
for the exchange part of the  self-energy (in the Matsubara representation):
\begin{equation}
\label{sf}
\Sigma_{\rm F}(\varepsilon_n) = 
4 g^2 \sigma i \varepsilon_n \ln\left( \frac{\rm  e}{12 \pi \sigma |\varepsilon_n| \tau} \right),
\end{equation}
where ${\rm e} \approx 2.718$ is the base of the natural logarithm [Eq.~(\ref{sf}) has better than
logarithmic accuracy]. From Eq.~(\ref{sf}), it follows that 
the negative exchange correction to the density of states is at most
$$
\delta \nu_{\rm F} (0,T) \propto - \nu g^2  \left(T\tau\right) \ln\frac{1}{T\tau}
$$
and does not contain the large factor $L$. Therefore, we conclude that the exchange correction to the
density of states is negligible in the low-temperature limit. 

A few remarks are in order: In the usual theory of disordered electronic systems it is well
known that the Hartree correction is parametrically smaller than the exchange correction in the case of
weak long-range interactions. We argue that there is no contradiction between this statement and the above results. 
In the context of two-dimensional electronic systems, a ``long-range interaction''
always implies the {\em screened} Coulomb interaction 
$\nu U(0,q) \sim r_{\rm s} \left[ q/(2 p_{\rm F}) + r_{\rm s}\right]^{-1}$. In real space, 
the screened Coulomb interaction falls off as $U_{\rm screened}(r) = e^2 a_{\rm B}^2/r^3$, where
$a_{\rm B}$ is the Bohr's radius.
The corresponding Hartree correction is finite, because the
interaction between an electron and density fluctuations far  from  it
does not diverge. For the Coulomb interaction, one can not
``control'' the overall strength and the screening properties independently. The limit of weak
 screening leads to the small value of the interaction itself. The interaction  averaged over the Fermi surface is
$\nu \left\langle U(0,{\bf p} - {\bf p}') \right\rangle \sim r_{\rm s} \ln(1/r_{\rm s}) \ll 1$.
Therefore the Hartree terms are small, while the Fock terms remain finite. In our theory,
the fermion-gauge-field coupling $g$ and the ``screening length'' $Q_{\rm min}^{-1}$ are completely
different parameters, and therefore the regime $Q_{\rm min}^{-1} \to \infty$ does not
imply an equally small $g$. In this limit, the infrared divergence of the Hartree diagram is not compensated
by the overall weakness of the interaction.

Summarizing this part, we present the correction to the density of states in leading order perturbation theory:
\begin{eqnarray}
\label{nu4}
\delta \nu (T) = \nu  \frac{g^2 L}{4 \pi^3 \delta} \ln\frac{1}{T\tau},
\end{eqnarray} 
where $g \ll 1$ is the fermion-gauge-field coupling, $\delta = \chi m$ with $\chi$ being the fermion ``magnetic susceptibility,''
and $L = \ln{\left( p_{\rm F}/Q_{\rm min} \right)}$, with $Q_{\rm min}^{-1}$ being a phenomenological cut-off parameter at this stage.  
For perturbation theory to be valid, we must require $| g^2 L \ln(T\tau) | \ll 1$. This is a condition on
the coupling $g$ and on temperature.

\subsection{Interaction corrections to conductivity}
\label{Sub:sigma}
In a non-interacting system, the conductivity is connected to the density of states via the simple 
 Einstein relation, $\sigma = e^2 \nu D$, where $D$ is the diffusion coefficient. In the
presence of interactions, the relation between the density of states and conductivity may be more 
complicated.
In this section, we calculate the Hartree and Fock corrections to conductivity using the
usual diagrammatic technique. 
\begin{figure}[htbp]
\centering
\includegraphics[width=3in]{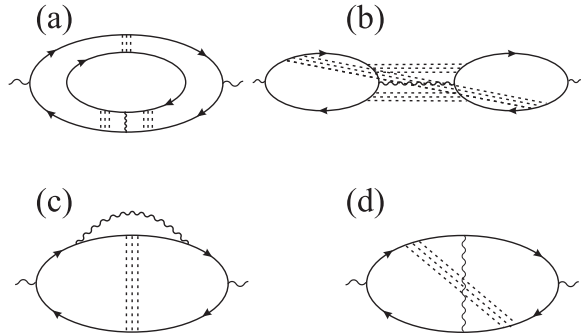}
\caption{Diagrams for the correction to conductivity. Graphs (a) and (b) are the Hartree contributions;
graphs (c) and (d) are the exchange (Fock) contributions. }
\label{Conductivity}
\end{figure}

The Hartree contributions to conductivity are given by diagrams (a) and (b) in Fig.~\ref{Conductivity}.~\cite{Fukuyama} 
It is possible to reduce the problem of
calculating these diagrams to the standard Altshuler-Aronov theory by simply redrawing the 
graphs as shown in Fig.~\ref{Saw-diagrams}, where the sawtooth line represents the effective interaction
(see also Fig.~\ref{Saw-diagrams}c):
\begin{equation}
\label{U}
\tilde{U}(\Omega,{\bf Q}) =  - \frac{2 B(\Omega,{\bf Q})} {\left( 2 \pi \nu \tau \right)^2}
\end{equation}
where $B$ is given by Eq.~(\ref{B}) with the fermionic frequencies and momenta set to zero.
In leading order perturbation theory, this function can be replaced by a constant (\ref{B0}). 
Therefore, the effective density-density interaction has essentially zero range, and the
usual theory applies. We note  that the scheme used here is possible only if one
can neglect the contribution coming from the explicit dependence of the gauge interaction on the 
external frequency $\omega$, which ``runs'' through the wavy line in Fig.~\ref{Conductivity}b. 
The corresponding contribution is small if 
$$
\frac{\sigma T}{\chi Q_{\rm min}^2} \ll  \ln\frac{1}{T\tau}
$$
or if one is interested in the AC conductivity with frequencies small but finite: $|\omega| > \chi Q_{\rm min}^2/\sigma$.
Here we assume that one of these conditions applies. In this case, we can use the following
formula for the correction to  conductivity
\begin{eqnarray}
\label{con_AA}
\nonumber
\delta\sigma^*_{\rm H} = i e_*^2 \frac{\sigma}{\pi} 
&& \!\!\!\!\! \int\limits_{-\infty}^{\infty} d\Omega \frac{\partial}{\partial \Omega}
\left[ \Omega \coth\left(\frac{\Omega}{2T}\right) \right]\\
&& \!\!\!\!\! \times
\int \frac{d^2{\bf Q}}{\left( 2 \pi \right)^2}
\frac{\tilde{U} DQ^2}{\left(-i \Omega +  DQ^2\right)^3},
\end{eqnarray}
where the effective interaction $\tilde{U}$ is given by Eq.~(\ref{U}) with $\Omega = 0$ and ${\bf Q} = {\bf 0}$.
We see that the Hartree contribution to  conductivity due to the long-range interaction $K(\Omega,{\bf Q})$
is described by the exchange contribution due to the short-range effective interaction $\tilde{U}$. Calculating
the integral in Eq.~(\ref{con_AA}), we obtain the following expression for the  conductivity:
\begin{figure}[htbp]
\centering
\includegraphics[width=3in]{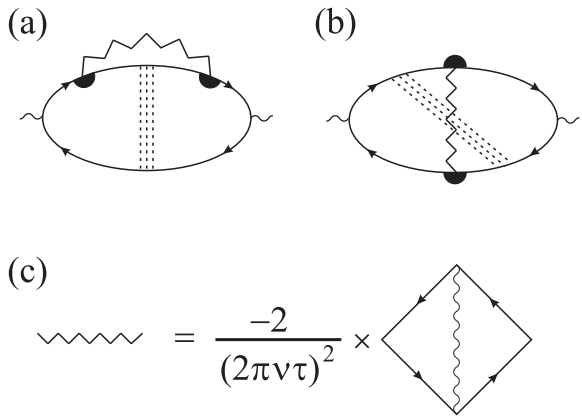}
\caption{(a) and (b):~Hartree diagrams expressed through the effective interaction $\tilde{U}$. 
(c):~The effective short-range interaction $\tilde{U}$ [see also Eq.~(\protect\ref{U})].}
\label{Saw-diagrams}
\end{figure}
\begin{equation}
\label{sigma_H}
\delta\sigma^*_{\rm H} =  \frac{e_*^2 g^2 \sigma L}{4 \pi^3 \delta} \ln\frac{1}{T\tau},
\end{equation}
where $e_*$ is the coupling (charge) to the field with respect to which we are calculating conductivity. 
Incidentally, we find that the Einstein relation holds for the Hartree contribution
to conductivity [see Eq.~(\ref{nu4})], $\delta\sigma^*_{\rm H} = e_*^2 \delta\nu_{\rm H} D$.

We now discuss the Fock contribution to the conductivity. It is given by graphs (c) and (d) 
in Fig.~\ref{Conductivity}. These diagrams were considered in Refs.~[\onlinecite{Khveshchenko,Mirlin_Wolfle}].
We rederive the corresponding contribution in Appendix~A, confirming the numerical
coefficient found in Ref.~[\onlinecite{Mirlin_Wolfle}]. The result is 
\begin{equation}
\label{sigma_F}
\delta\sigma^*_{\rm F} = -\frac{e_*^2 g^2 \ln{\sigma}}{\pi^2} \ln\frac{1}{T\tau}.
\end{equation}
We see that the Einstein relation does not apply for the Fock contribution.
The Fock correction to conductivity is relatively large, eventhough the 
Fock correction to the density of states is essentially zero. Nevertheless,
the exchange correction (\ref{sigma_F}) is still much smaller than the Hartree 
term (\ref{sigma_H}) due to the singularity in the factor $L$.

\section{Conclusion}
\label{Sec-Conclusion}

In this paper we found that in a two-dimensional Fermi system with singular interactions,
the dominant quantum corrections to the density of states and conductivity are positive,
which suggests that the system  remains in a metallic state at low temperatures. By singular interactions we mean
long-range interactions $V(\Omega,{\bf Q})$ for which the integral $L = \int d^2 {\bf Q} V(0,{\bf Q})$
is infrared divergent.

Singular current-current interactions may appear in the context of quantum magnets~\cite{Ioffe_Larkin}
and the quantized Hall effect,~\cite{HLR} and correspond to the interactions mediated by {\em transverse}
gauge fields.  If the fermion-gauge-field coupling is small (large-$N$  limit),
 a controllable theory can be developed. In the clean case, the system is in the 
strong coupling phase discovered in Refs.~[\onlinecite{Nayak_Wilczek,AIM,Polchinski}]. This non-Fermi liquid phase
is described by a collection of regions on the Fermi line, which do not
interact with each other, unless they are located on  opposite sides of the Fermi line.
In the latter case, the regions interact strongly, and the strength of these $2p_{\rm F}$-interactions
is non-universal, in the sense that it depends on the effective gauge coupling. These
interactions may lead to instabilities at low temperatures.~\cite{AILM,VG_tbp}     
At high enough temperatures or in the absence of instabilities, the system remains gapless.
This gapless non-Fermi liquid phase is unstable against weak disorder. At low but finite temperatures 
and in the presence of disorder, the fermion-gauge system is described by a diffusive Fermi liquid
coupled to a ``dissipative'' gauge field. This ``dissipative'' gauge field remains singular
({\em i.e.}, long-range) in the sense defined above. This {\em transverse} gauge field has two important effects on electron
transport: (i)~First, gauge field fluctuations suppress the weak localization effect by inducing fluxes through closed
diffusive trajectories and lead to dephasing topological in origin. At finite temperatures,
classical fluctuations of magnetic field lead to a relatively large dephasing rate~\cite{Aronov_Wolfle} proportional to conductance
rather than resistance ({\em c.f.}, Ref.~[\onlinecite{Altshuler_Aronov}]). It is conceivable that similar quasistatic configurations due
to quantum fluctuations of the magnetic field still result in dephasing. The corresponding time-scale (\ref{tau_*})
should be related to the flux-flux correlator calculated in this paper, which was found to be proportional to the
perimeter of the trajectory.
(ii)~Second, gauge fluctuations result in singular interaction corrections to conductivity. 
Due to interactions being long-range, the main contributions come from the processes describing
scattering of fermions off of the static self-consistent potential created by the other fermions. 
The corresponding Hartree contribution to conductivity is positive and formally infrared divergent.  
This strongly suggests that the metallic phase is stable. 

We note that somewhat similar singular interactions may appear in electronic systems in the vicinity of quantum
critical points. In this case, the role of a singular {\em  longitudinal} gauge field is played
by soft fluctuation modes.~\cite{referee} In the clean case,  treatment of these fluctuations is  similar 
to the theory developed in Refs.~[\onlinecite{Nayak_Wilczek,AIM,Polchinski}]. 
In particular,  Eliashberg-type theories developed to describe Fermi systems in the vicinity
of quantum critical points~\cite{VG_Yakovenko,Locality} are  closely related to the strong-coupling theory of Nayak and Wilczek
and Altshuler {\em et al}. It also seems  likely that the non-Fermi liquid phase discovered in the context of gauge
theories can be realized in the vicinity of quantum critical points. If the transition 
happens at zero wave-vector $q_0 = 0$, then a direct mapping exists between the two theories. 
 If the transition happens at a non-zero wave vector~\cite{VG_Yakovenko} $|{\bf q}| = q_0 \leqslant 2 p_{\rm F}$, then 
the existence and structure of the non-Fermi liquid phase should depend on
whether $z = \arcsin{\left( q_0/2p_{\rm F}\right)}/\pi$ is a rational number or not. If $z=p/q$ is rational, then   
the system is described  by a collection of sets of regions on the Fermi line ($q$ regions in each set),
with different sets not interacting with each other, unless the momentum transfer between two sets is exactly 
two Fermi-momenta. These strong-coupling non-Fermi liquid fixed points,
if exist, are unstable against weak disorder. In the diffusive regime, singular interactions should
modify quantum corrections to conductivity. It is possible that in certain cases long-range interactions
mediated by soft fluctuation modes may enhance the Hartree terms compared to the other contributions.
In these cases,  the competition between negative weak localization
and exchange corrections to conductivity and positive Hartree terms  would result in a quantum
metal-insulator transition.

A serious limitation of the approach used here and in 
Refs.~[\onlinecite{Nayak_Wilczek,AIM,Polchinski}] is the assumption
about a weak gauge-field-fermion coupling. In the systems
of interest, there are no physical reasons for this coupling to be small.
We note that it however might be possible to connect the results derived
within the large-$N$ approximation to real systems. This can be done
by relating the effective coupling strength to observables:  
In the clean strong coupling  phase, the Kohn anomaly in the fermion polarizability
can be much stronger~\cite{AIM,IM} than in   an ordinary Fermi liquid.~\cite{Stern} The corresponding singularity 
is related to the strength of the gauge-field-fermion coupling. One can calculate the general
form of the dynamic polarizability near $q = 2 p_{\rm F}$,~\cite{VG_tbp} which then determines 
unusual dynamic Friedel oscillations. It is known that in an ordinary Fermi liquid, scattering of electrons
off of the dynamic Friedel oscillations leads to a ``non-analytic'' linear-in-$T$ temperature dependence of
thermodynamic properties~\cite{VG_SDS_m(T),Maslov_Chubukov,Chubukov_etal,VG_Chubukov} and transport 
properties~\cite{sds-T,Zala_etal} in the ballistic  temperature regime
(where ``non-analytic'' means that these corrections are different from the $T^2$-corrections expected in the
na{\"{\i}}ve Fermi gas theory). In the fermion-gauge theory, these corrections scale differently with the exponent
being related to the effective gauge-field-coupling.~\cite{VG_tbp} Therefore by measuring these non-universal temperature corrections ({\em e.g.}, 
corrections to the specific heat, susceptibility, electrical transport, and thermal transport), one in principle can determine the effective
gauge coupling. This construction relies on the assumption  that the strong-coupling phase
survives at intermediate couplings $g \sim 1$, which is a reasonable conjecture.

\begin{acknowledgments}
The author is grateful to Leon Balents, Matthew  Fisher, Pavel Kovtun,
Anatoly Larkin, Andreas Ludwig, Lesik Motrunich, Gil Refael, and Maxim Vavilov for discussions about this work. 
This work was supported by the  David and Lucile Packard foundation. 
\end{acknowledgments}

\appendix
\section{Exchange Correction to Conductivity}
\label{sec-ap}

In this Appendix we derive the formula for the exchange correction to conductivity.
The leading exchange contribution to the conductivity is coming from diagrams (c) and (d) in Fig.~\ref{Conductivity}.
It is not {\em a priori} evident that one can use the formulas from the standard Altshuler-Aronov theory of density-density
interaction corrections; since in the latter case, the scalar vertices are renormalized by disorder,
which produces additional constraints on the frequency summations. We do not have these restrictions.
Nevertheless, we show below that the analytical structure of the exchange contributions in our case 
is  essentially identical to the structure of the standard Altshuler-Aronov treatment.

The contributions to the ${\bf A}$-field response  tensor from diagrams (c) and (d) are
\begin{eqnarray}
\label{Qd}
\nonumber 
{\cal Q}^{\rm (c)}_{\alpha\beta}(\omega) = {e_*}^2 g^2 T \sum\limits_\varepsilon \int &&\!\!\!\!\!\! 
\frac{d^2{\bf Q}} {\left( 2 \pi \right)^2}
 T_{\alpha\alpha'} (\varepsilon,\omega,\omega+\Omega) \\
&&\!\!\!\! \hspace*{-1.75in} \times\, 
T_{\beta\beta'} (\varepsilon,\omega,\omega+\Omega) K_{\alpha'\beta'} (\Omega,{\bf Q}) {\cal D}_{\varepsilon,\varepsilon+\omega+\Omega} ({\bf p},{\bf p + Q})
\end{eqnarray}
and
\begin{eqnarray}
\label{Qe}
\nonumber 
{\cal Q}^{\rm (d)}_{\alpha\beta}(\omega) = {e_*}^2 g^2 T \sum\limits_\varepsilon \int &&\!\!\!\!\!\! \frac{d^2{\bf Q}} {\left( 2 \pi \right)^2}
 T_{\alpha\alpha'} (\varepsilon,\omega,\omega+\Omega) \\
&&\!\!\!\!  \hspace*{-1.75in}  \times\, T_{\beta\beta'} (\varepsilon,\Omega,\omega+\Omega)
K_{\alpha'\beta'} (\Omega,{\bf Q}) {\cal D}_{\varepsilon,\varepsilon+\omega+\Omega} ({\bf p},{\bf p + Q}),
\end{eqnarray}
where functions $T$ correspond to the following block of three fermionic Green's functions
\begin{eqnarray}
\label{GGG}
\nonumber
T_{\alpha\beta}(\varepsilon,\omega_1,\omega_2) = \int&& \!\!\!\!\!\!  \frac{d^2{\bf p}} {\left( 2 \pi \right)^2}
\frac{p_\alpha}{m} \frac{p_\beta}{m}
{\cal G}_{\varepsilon} ({\bf p}) {\cal G}_{\varepsilon+\omega_2} ({\bf p}) {\cal G}_{\varepsilon+\omega_1} ({\bf p})\\ 
&& \!\!\!\!\!\! \times \,\theta\left[-\varepsilon \left(\varepsilon+\omega_2\right)\right].
\end{eqnarray}
Using the free fermion Green's functions~(\ref{G0}), we find:
\begin{eqnarray}
\label{T}
\nonumber
T_{\alpha\beta}(\varepsilon,\omega_1,\omega_2) =&& \!\!\!\!\!\! 2 \pi \sigma \tau {\rm sgn} (\varepsilon)
\theta\left[-\varepsilon \left(\varepsilon+\omega_2\right)\right] \\
&& \!\!\!\!\!\! \hspace*{-0.5in} \times\, \left\{ \theta\left[-\varepsilon \left(\varepsilon + \omega_1\right)\right] -
\theta\left[\varepsilon \left(\varepsilon+\omega_1\right)\right] \right\}.
\end{eqnarray}
The last factor is introduced to take into account the constraint on the frequency sum originating 
from the diffuson (with $\omega_2 =\omega+\Omega$).
Using Eqs.~(\ref{Qd}), (\ref{Qe}), and (\ref{T}), we can write the total response tensor 
due to two pairs of diagrams of type (c) and (d). This sum has the same analytical
structure as the corresponding sum in the usual Altshuler-Aronov theory  even though they
set of $\theta$-functions involved is slightly different.  
\begin{eqnarray}
\label{Qd+Qe}
\nonumber
{\cal Q}_{\alpha\beta} (\omega) = && \!\!\!\!\! 2 {\cal Q}^{\rm (c)}_{\alpha\beta}(\omega)+
2 {\cal Q}^{\rm (d)}_{\alpha\beta}(\omega) \\
&& \!\!\!\!\!\!\!\!\!\! \hspace*{-0.4in} =- 4 \sigma D T 
\left[ \sum\limits_{\Omega=0}^{\omega} \Omega F(\omega,\Omega) + 
\sum\limits_{\Omega=\omega}^{\infty} \omega F(\omega,\Omega) \right],
\end{eqnarray}
where 
\begin{equation}
\label{F}
F(\Omega,\omega) = \int \frac{d^2{\bf Q}} {\left( 2 \pi \right)^2}
\frac{1}{|\Omega + \omega| + DQ^2}
\frac{1}{\sigma |\Omega| + \chi Q^2}.
\end{equation}  
Equations~(\ref{Qd+Qe}) and (\ref{F}) were first derived in Ref.~[\onlinecite{Mirlin_Wolfle}]
by Mirlin and W{\"o}lfle. 
Using the standard analytical continuation procedure
(see Appendix~4. of Ref.~[\onlinecite{Altshuler_Aronov}] for details), they found the correction to the DC conductivity as
\begin{eqnarray}
\label{sigmaAA_0}
\nonumber
\delta\sigma^*_{\rm F} = - \frac{ {e_*}^2 g^2 \sigma D}{\pi}
{\rm Im}\, \int && \!\!\!\! d\Omega \frac{\partial}{\partial \Omega} 
\left[ \Omega \coth\left(\frac{\Omega}{2T}\right) \right]\\
&& \!\!\!\! \times F(0,|\Omega| \to -i\Omega).
\end{eqnarray}
At low temperatures, this  leads to the following result:
\begin{equation}
\label{sigmaAA_01}
\delta\sigma^*_{\rm F} = - \frac{{e_*}^2 g^2 \ln{\sigma}}{\pi^2} \ln{\frac{1}{T\tau}}. 
\end{equation}

\bibliography{gauge}

\end{document}